\documentclass[11pt]{article}

\usepackage[latin1]{inputenc}
\usepackage[OT1]{fontenc}
\usepackage[english]{babel}
\usepackage{fullpage}
\usepackage{amsmath}
\usepackage{amssymb}
\usepackage{amsthm}
\usepackage{mathrsfs}
\usepackage{graphicx}
\usepackage{dsfont}
\usepackage{wrapfig}
\usepackage{comment}
\usepackage{rotating}
\usepackage{hyperref}
\usepackage{subfigure}
\usepackage{algorithmic}
\usepackage[ruled,linesnumbered,shortend,vlined,noend]{algorithm2e}
\usepackage[all]{xy}
\usepackage{authblk}

\newcommand{\R}{\mathbb{R}}

\newtheorem{thm}{Theorem}[section]

\newtheorem{defin}[thm]{Definition}

\title{Topological Data Analysis of Single-cell Hi-C Contact Maps}
\author[1]{Mathieu Carri\`ere}
\author[1]{Ra\'ul Rabad\'an}
\affil[1]{Department of Systems Biology, Department of Biomedical Informatics, Columbia University, New York, US. }
\begin{document}

\maketitle

\begin{abstract}
In this article, we show how the recent statistical techniques developed in Topological Data Analysis for the Mapper algorithm
can be extended and leveraged to formally define and statistically quantify the presence of topological structures coming from
biological phenomena in datasets of CCC contact maps.
\end{abstract}

\section{Introduction}

The three dimensional structure of chromosomes varies across cell types, cell states, and the cell cycle. Recent technological developments allow us to answer 
fundamental questions about this three dimensional structure and its relationship to other biological processes. For instance, how chromatin folds at different scales,  
how chromatin states change during dynamical biological processes such as differentiation and the cell cycle, how to compare chromatin states between healthy and diseased cells, 
and how chromatin states relate to other biological processes. It is known that chromosomes tend to fold up tightly in preparation for mitosis and 
rapidly unfold once mitosis is finished in order to enable several functions such as DNA transcription and replication. Studying and mathematically 
quantifying the evolution of chromosomal spatial organization would thus allow for a better understanding of the epigenetic dynamics of the cell.
A common way to study the spatial conformation of DNA is through chromosome conformation capture (CCC)~\cite{Dekker02, Wit12}, which is a set of techniques tracking
the regions of the genome that are in close spatial proximity. Although the first methods that were chronologically developed in CCC, such as 4C or 5C methods~\cite{Simonis06, Dostie06}, 
were limited in the number of genomic regions that one can observe, recent technological breakthroughs in high-throughput sequencing have enabled to retrieve 
the spatial proximities between genomic regions at the scale of the whole genome~\cite{Aiden09}. This genome-wide CCC technique is usually denominated by Hi-C, 
and encodes these proximities in a contact matrix, or contact map, whose rows and columns represent bins of the genome at a specific genomic resolution. Since these technological 
breakthroughs have lead to the generation of large datasets of such matrices, the question of an efficient analysis method has become of primary importance. A few methods 
and similarity functions have been proposed within the last few years, but general methods with high statistical power and guarantees are still missing in the CCC literature.
Topological Data Analysis (TDA)~\cite{Carlsson09a} has been emerging over the last decade as one possible answer to this problem. 
Indeed, it is a very general and mathematically-grounded statistical set of tools building on topology, which enjoys several useful guarantees, 
such as robustness and invariance to solid deformations of data~\cite{Cohen07}.

\paragraph*{Contributions.} 
In this article, we present a formal way to efficiently encode and statistically assess the presence of structure in a dataset of 
CCC contact maps using TDA. We then apply this method to a Hi-C contact map dataset, and show that we are able to successfully retrieve, 
encode and quantify the biological information that was empirically observed in this data.

\section{CCC and Hi-C contact maps}

\subsection{Background}

The 3D spatial organization of chromosomes is measured with a set of methods belonging to chromosome conformation capture (CCC or 3C)~\cite{Wit12}. 
This family of protocols detects the loci of the genome that are spatially close and interacting in the nucleus of the cell. There exists several different methods, 
each of which captures a different level of chromosomal interactions: the original 3C experiments~\cite{Dekker02} quantify the interaction between a given pair of loci of the genome, 
4C experiments~\cite{Simonis06} allow one to measure the interaction between a given locus and all others loci and 5C experiments~\cite{Dostie06} measure the interactions 
between all pairs of loci in a specific region. More recently, the development of high-throughput sequencing technologies has enabled the use of Hi-C experiments~\cite{Aiden09}, 
where interactions between all pairs of loci are quantified  genome-wide. Moreover, the use of Hi-C experiments at single-cell resolution has just begun and is one 
of the most exciting research topics in genomics.
  
\begin{figure}[h]\centering
\includegraphics[width = 15cm]{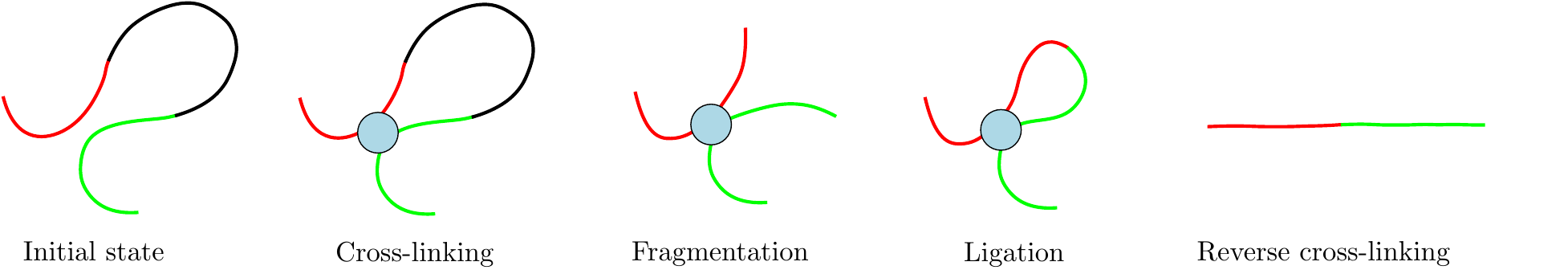}
\caption{\label{fig:CCC} Scheme of the four steps of the CCC procedure.}
\end{figure}

Even though these methods usually differ in their final steps, they all begin with the same first four steps. First, pairs of loci that are spatially close 
are cross-linked with formaldehyde, so as to create a robust bond between the loci. Then, the chromatin is cut and fragmented with a restriction enzyme, the fragment length 
specifying the resolution of the experiment. In the third step, the fragment pairs are ligated through the action of an enzyme, thus forming loops, that are eventually broken 
by reverse cross-linking. At the end of this fourth step, the data is comprised of a large set of isolated fragment pairs. See Figure~\ref{fig:CCC} for an illustration.

In practice, these fragment pairs have to be amplified to be detectable, and sequenced to a reference genome in order to retrieve the loci to which 
they belong to. This is generally where the previous methods differ, using protocols ranging from simple polymerase chain reaction (PCR) amplification, in 
which a DNA polymerase enzyme is used to increase the fragment concentration, to more refined additional ligation and enrichment cycles of the fragment before sequencing.
 
\begin{figure}[h]\centering
\includegraphics[width = 15cm]{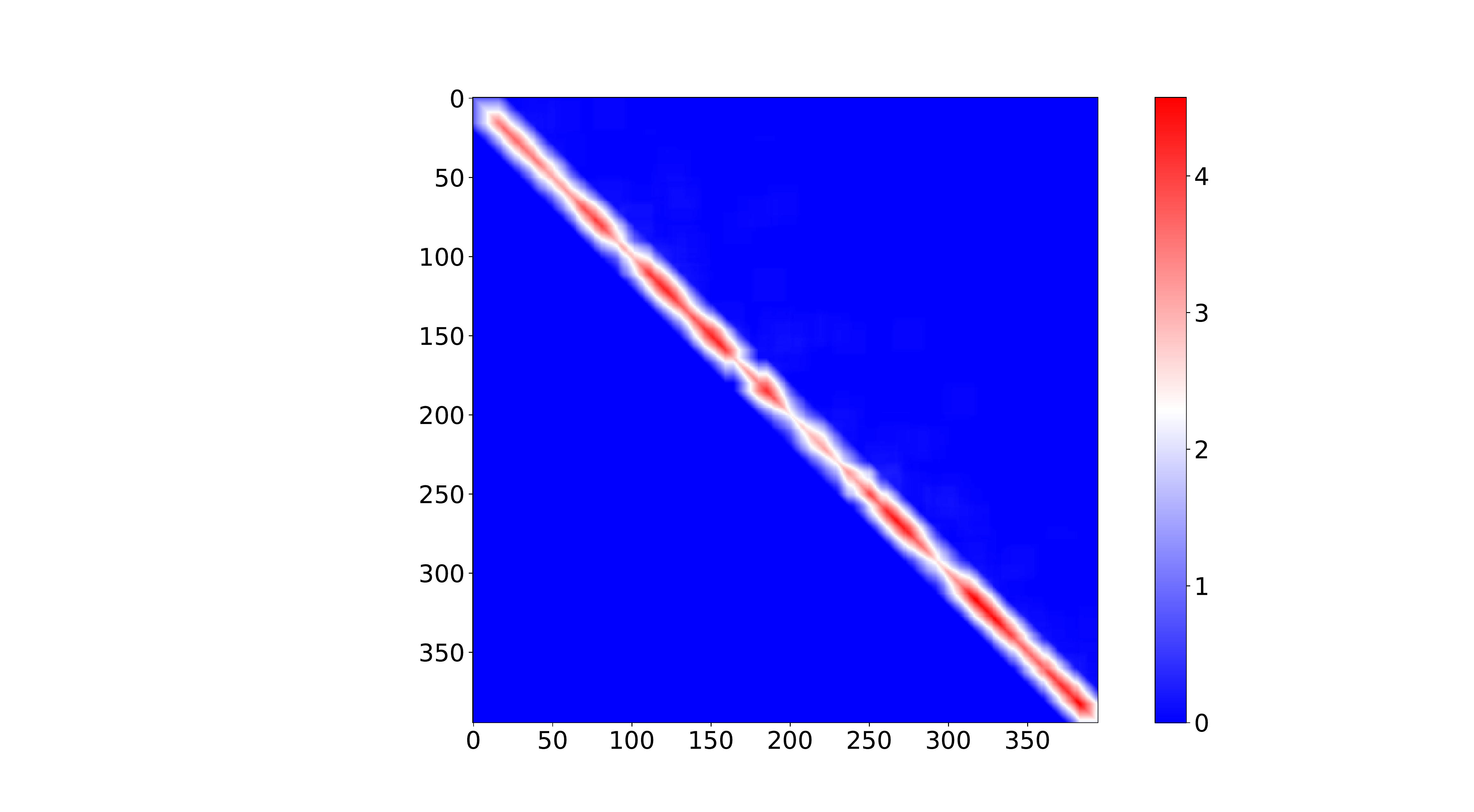}
\caption{\label{fig:map} Example of contact map on a single chromosome with 500kb bins.}
\end{figure}

For all these methods, the final information is encoded in a so-called {\em contact map}, which is a symmetric matrix, whose rows and columns represent small loci, or bins, 
of the genome. The entry in position $(i,j)$ of this matrix is an integer equal to the number of pairs whose fragments belong to bin $i$ and bin $j$. These matrices are usually very large, 
their sizes depending directly on the bin resolutions, and sparse, with larger positive values on the diagonal. See Figure~\ref{fig:map} for an example of such a matrix.

\subsection{Preprocessing and comparison procedures}

It is now known that the previous methods suffer from various biases, and that these matrices require significant preprocessing in order to be analyzed. There is a large 
literature on possible preprocessing methods. See~\cite{Ay15} for a comprehensive review. However, it has been shown recently that the so-called 
{\em stratum-adjusted correlation coefficient} ${\rm SCC}$ is an efficient and powerful quantity for comparing Hi-C matrices.

This coefficient is inspired from the generalized Cochran-Mantel-Haenszel statistic~\cite{Mantel63,Agresti12}, which is used to understand the correlation between two variables that are both stratified by a third one.
Roughy, the SCC is computed by grouping the entries of the Hi-C matrices with respect to the distance between the corresponding loci of each entry, and to compute a 
weighted average of the corresponding Pearson correlations.

\begin{defin}{\rm \bf Stratum-adjusted correlation coefficient~\cite{Yang17}.} Let $X,Y\in\mathcal S_n$ be two Hi-C matrices with $n$ bins. For each $1\leq k\leq n$, let $N_k=\{ (i,j)\,:\,1\leq i,j \leq n \text{ and } k-1 < |j-i| \leq k \}$
be the set of indices which represent fragments separated by $k$ bins, and let $X_k=\{X_{i,j}\,:\, (i,j)\in N_k\}$ and $Y_k=\{Y_{i,j}\,:\, (i,j)\in N_k\}$. Then, the 
stratum-adjusted correlation coefficient {\rm SCC} is defined as:
\begin{equation}
{\rm SCC}(X,Y) = \frac{\sum_k {\rm card}(N_k) {\rm Cov}(X_k,Y_k) }{\sum_k {\rm card}(N_k) \sqrt{{\rm Var}(X_k){\rm Var}(Y_k)}}
\end{equation}
\end{defin}

As noted by the authors in~\cite{Liu18}, the SCC provides a useful similarity function for Hi-C contact maps. Indeed, using it in conjunction with classical dimensionality reduction techniques, 
such as PCA or MDS, on the recent single-cell Hi-C dataset of~\cite{Nagano17}, allowed them to successfully group cells according to the phase of the cell cycle they belonged to. However, 
there is still a lack of formalization when it comes to the analysis. Indeed, the authors in~\cite{Liu18} note that the MDS embedding obtained with SCC has a circular shape which roughly 
reflects the cell cycle, but there is no mathematical formalization or statistical guarantee of this observation. On the other hand, in this work, we use the SCC to build a topological 
representation of the Hi-C data that formally demonstrates the existence of an intrinsic circular shape in the data with high confidence, and which efficiently encodes the biological information 
corresponding to this pattern.

\section{Topological Data Analysis and Mapper}

\subsection{Background}

Topological Data Analysis (TDA) is a growing method in the field of data science, whose main goal is to extract and encode the topological information contained in geometric data.
It has been shown in many cases that this kind of information can be particularly relevant to data analysis, and often improves results when combined with other traditional statistical descriptors.
TDA has now found its way into many applications, and has encountered significant success in genomics when applied to 
single-cell RNA sequencing data~\cite{Camara16, Rizvi17}.

TDA is usually carried through the computation of descriptors coming from the analysis of simplicial complexes built on datasets. 
One of such descriptors is the so-called {\em Mapper} simplicial complex,
whose topology can be shown to capture the underlying topology of the data.
This complex requires a {\em filter function}, sometimes called {\em lens}, defined on the data. Roughly,
the Mapper is computed by first covering the image of the filter with overlapping hypercubes, and then taking the nerve of 
the connected components of each preimage of these hypercubes.

\begin{defin}{\rm \bf Mapper~\cite{Singh07}.} Let $X\subset\R^d$, and $f:X\rightarrow\R^p$ be a function defined on $X$. Let 
$\mathcal U=\{U_i\}$ be a cover of ${\rm im}(f)$, i.e. a family of sets such that ${\rm im}(f)\subseteq \cup_i U_i$, let $\mathcal V = f^{-1}(\mathcal U) = \{f^{-1}(U_i)\}$
be the cover obtained by taking the preimage of $\mathcal U$ under $f$, and let $\tilde{\mathcal V} = \pi_0(\mathcal V)$, i.e. the cover obtained by 
separating each element of $\mathcal V$ into its connected components. 
Then, the {\em Mapper} ${\rm M}_f(X,\mathcal U)$ is defined as:
\begin{equation}
{\rm M}_f(X,\mathcal U) = \mathcal N (\tilde{\mathcal V}),
\end{equation} 
where $\mathcal N$ denotes the nerve operation.
\end{defin}

In practice, the input space $X$ is given as a point cloud, the connected components are derived with a clustering algorithm, and the pairwise intersections of elements of $\tilde{\mathcal V}$,
which are necessary to compute the nerve, are retrieved by detecting points shared by multiple elements of the cover. It is also very common to use covers with fixed-size hypercubes,
i.e. each $U_i$ is defined as a product of intervals $[a_1,b_1]\times\cdots\times[a_p,b_p]$, and to use the same interval length, or {\em resolution}, and overlap percentage, or {\em gain},
for each of the $p$ dimensions of the filter. 
We also emphasize that, even though the Mapper is defined on point clouds, the mere distance matrix between the points might be enough to compute it, 
as long as the clustering algorithm only requires the pairwise distances, which is the case for i.e. single-linkage clustering.

It is known that the Mapper, when computed on a point cloud $X$ sampled from an underlying object $\mathbb X$ such as a manifold, 
is actually nothing but an approximation of a limiting object, called the Reeb space ${\rm R}_f(\mathbb X)$, defined as the quotient space of $\mathbb X$
with the relation $\sim_f$:

\begin{equation}\label{eq:reeb}
{\rm R}_f(\mathbb X) = \mathbb X / \sim_f,
\end{equation}

where $\sim_f$ identifies points $x,y$ that satisfy $f(x)=f(y)$ and that belong to the same connected component of $f^{-1}(f(x)) = f^{-1}(f(y))$.
Hence, the natural question to ask is how close a Mapper is to its limiting Reeb space.
There exists several theoretical guarantees on this topic in the literature~\cite{Dey16, Munch16, Carriere17b}, showing different types and quantifications of convergence of the Mapper 
depending on the assumptions that are made on the computation of this complex. For instance, the authors in~\cite{Munch16} use a specific metric between Mappers to prove a very general type
of convergence to the limiting Reeb space. However, computation algorithms and interpretation of this metric are still lacking in the literature. On the other hand, 
the authors in~\cite{Carriere17b} chose to focus on Mappers computed with scalar-valued filters $f:X\rightarrow\R$ and single-linkage clustering with fixed threshold $\delta>0$.
In this simpler setting, comparing Mappers with their so-called {\em extended persistence diagrams} allowed them to derive precise bounds and information about the topology 
given by the Mapper.

\subsection{Statistics on 1-dimensional Mappers} 
\label{sec:stats}

In this section, we briefly present how {\em extended persistence diagrams}~\cite{Cohen09} are used to compute statistics on 1-dimensional Mappers. 
We refer the interested reader to~\cite{Carriere18a} for further details.
We start by defining the Mapper that is computed in practice.

\begin{defin}
Let $X\subset\R^d$ be a point cloud, and $f:X\rightarrow\R$ be a function defined on $X$. Let 
$\mathcal I=\{I_i\}$ be a cover of ${\rm im}(f)$, i.e. a family of intervals such that ${\rm im}(f)\subseteq \cup_i I_i$, let $\mathcal V = f^{-1}(\mathcal I) = \{f^{-1}(I_i)\}$
be the cover obtained by taking the preimage of $\mathcal I$ under $f$.
Let $\tilde{\mathcal V} = {\rm SL}_\delta(\mathcal V)$ be the cover obtained with
single-linkage clustering with parameter $\delta$ on each element of $\mathcal V$. 
Then, the {\em Mapper} ${\rm M}_{f,\delta}(X,\mathcal I)$ is defined as:
\begin{equation}
{\rm M}_{f,\delta}(X,\mathcal I) = \mathcal N (\tilde{\mathcal V}),
\end{equation} 
where $\mathcal N$ denotes the nerve, and where intersection is determined by the presence of common points.
\end{defin}

\paragraph*{Parameter selection.} In practice, if $X_n\subset\R^d$ is a point cloud with $n$ points randomly sampled from a compact submanifold $\mathbb X$ of $\R^d$,
techniques from statistical support estimation can be used to choose $\delta_n>0$ so as to make the $\delta_n$-neighborhood graph $G_{\delta_n}(X_n)$
a good estimate of $\mathbb X$. The authors of~\cite{Carriere18a} suggest using $\delta_n=d_{\rm H}(X_n,X_{s(n)})$, where $d_{\rm H}$ denotes the Hausdorff distance
and $X_{s(n)}$ is a subsampling of $X_n$ of cardinality 
\begin{equation}\label{eq:sn}
s(n)=\frac{n}{{\rm log}(n)^{1+\beta}}, 
\end{equation}
with $\beta>0$.
However, due to the approximation induced by the heuristic used to assess intersections of cover elements, it may happen that discretization artifacts lead to major differences
with the target Mapper ${\rm M}_f(G_{\delta_n}(X_n),\mathcal I)$, even if $G_{\delta_n}(X_n)$ correctly approximates the underlying support $\mathbb X$.

\begin{thm}{\rm \cite{Carriere17b}}\label{th:1dmap}
Let $X_n\subset\R^d$ be a point cloud, and
let $f:X_n\rightarrow\R$ be a function defined on $X_n$.
Let $\delta_n=d_{\rm H}(X_n,X_{s(n)})$, and let
$\mathcal I_n$ be a cover of ${\rm im}(f)$ with gain $g\in\left(\frac 13, \frac 12\right)$ and resolution $r = {\rm max}\{|f(x)-f(y)|\,:\,\|x-y\|\leq\delta\} / g$.
Then ${\rm M}_{f,\delta_n}(X_n,\mathcal I)$ and ${\rm M}_f(G_{\delta_n}(X_n),\mathcal I)$ are isomorphic.
\end{thm}


\paragraph*{Bootstrapping.} Even if the previous theorem gives a nice heuristic to compute optimal parameters, it does 
not provide confidence regions for the Mapper. A general way to obtain such confidence regions is with {\em bootstrapping}. However,
bootstrapping techniques require at least an easily computable metric between the considered statistics. This is why comparing Mappers with their
{\em extended persistence diagrams}~\cite{Cohen09}, for which we have such a computable metric, is a reasonable approach.
  
\begin{figure}[h]
\includegraphics[width = 15cm]{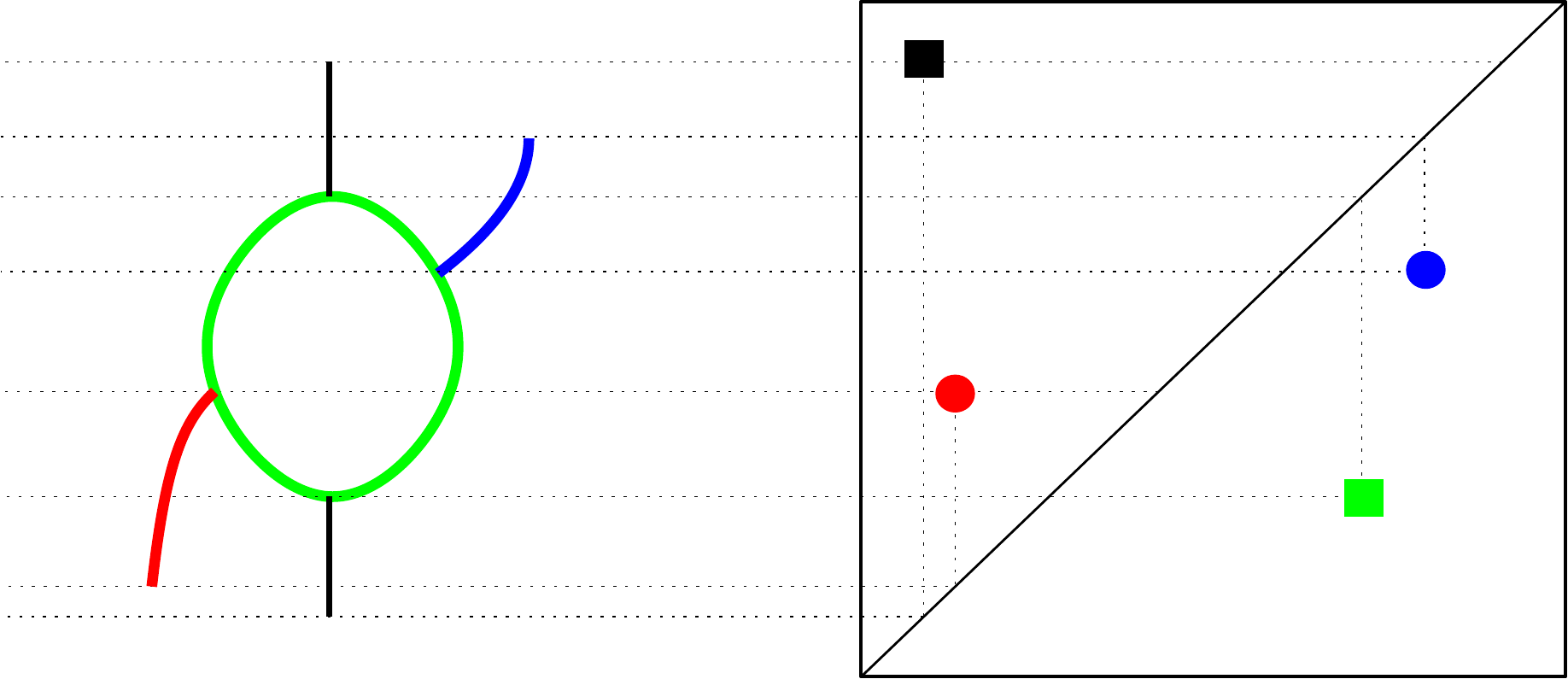}
\caption{\label{fig:topodict} Example of graph topology formalization with an extended persistence diagram. Each topological feature of the graph is mapped to a point
in the diagram, whose distance to the diagonal $\Delta$ characterizes the feature size.}
\end{figure}

We do not go into detail about extended persistent homology in this work, and we refer the interested reader to~\cite{Edelsbrunner10,Oudot15} for a thorough treatment of persistence theory.
We only recall that an extended persistence diagram requires a simplicial complex and a scalar-valued function $f$ defined on its nodes to be computed, and that it
takes the form of a set of points in the plane $\R^2$, such that each point represents a topological feature of the simplicial complex seen through the function $f$.   
Moreover, each point has a type specifying the topological feature it represents (connected component, cycle, cavity...), and
the distance of a point to the diagonal $\Delta=\{(x,x)\,:\,x\in\R\}\subset\R^2$ actually provides the size of the corresponding feature:
the farther away from the diagonal a point is, the bigger, or more significant, the corresponding topological feature.
Finally, note that contrary to ordinary persistence diagrams, points in extended persistence diagrams may also be located below the diagonal, due to the fact
that extended persistence is computed with both sub- and superlevel sets of the function $f$.
See Figure~\ref{fig:topodict} for an illustration.

What makes the use of extended persistence diagrams interesting is that they come equipped with a metric, the {\em bottleneck distance}, which is easily computable:

\begin{defin}
Given two extended persistence diagrams $D,D'$, a \emph{partial matching} between $D$ and $D'$ is a subset $\Gamma$ of $D\times D'$ such that:
$\forall p\in D$, there is at most one $p'\in D'$ such that $(p,p')\in\Gamma$, and 
$\forall p'\in D'$, there is at most one $p\in D$ such that $(p,p')\in\Gamma$. 
The \emph{cost} of $\Gamma$ is:
\[{\rm cost}(\Gamma)=\max\left\{\max_{p\in D}\ \delta_D(p),\ \max_{p'\in D'}\ \delta_{D'}(p')\right\},\]
where $\delta_D(p)=\|p-p'\|_\infty$ if $\exists p'\in D'$ such that $(p,p')\in\Gamma$, otherwise $\delta_D(p)=\inf_{q\in\Delta} \|p-q\|_\infty$,
and $\delta_{D'}(p')=\|p-p'\|_\infty$ if $\exists p\in D$ such that $(p,p')\in\Gamma$, otherwise $\delta_{D'}(p')=\inf_{q\in\Delta} \|p'-q\|_\infty$.
The \emph{bottleneck distance} between $D$ and $D'$ is then defined as:
\[d_{\rm b}(D,D')=\inf_{\Gamma}\ \emph{cost}(\Gamma),\]
where $\Gamma$ ranges over all partial matchings between $D$ and $D'$.
\end{defin}

This distance can be used to compute confidence intervals on the Mapper since one can bootstrap the point cloud in order to generate a distribution of bottleneck distances.
This distribution can then be used to assess confidence and compute p-values for each point of the diagram, and thus on the corresponding features in the Mapper as well. 
A typical scheme is the following one:

\begin{itemize}
\item Draw $X^1_n,\cdots,X_n^N$ from $X_n$ with replacement
\item Compute all $d_i = d_{\rm b}({\rm M}_{f,\delta_n}(X_n,\mathcal I_n),  {\rm M}_{f,\delta_n}(X^i_n,\mathcal I_n))$
\item Approximate $\mathbb P(d_{\rm b}({\rm M}_{f,\delta_n}(X_n,\mathcal I_n), {\rm R}_f(\mathbb X)) \leq\alpha)$ with $\frac {1}{N} {\rm card}(\{i\,:\,d_i\leq\alpha\})$ 
\end{itemize}

Hence, given a size $\alpha > 0$,  the previous procedure provides a way to compute the confidence level of which all features of size at least $\alpha$ are actually
also present in the limiting Reeb space ${\rm R}_f(\mathbb X)$ as defined in Equation~(\ref{eq:reeb}), and reciprocally, given a confidence $c\in[0,1]$, it also provides a way to visualize 
the features on which we have confidence at least $c$. Indeed, for each confidence $c$, the procedure outputs a value of the bottleneck distance $d_c$ such that
$\mathbb P(d_{\rm b}({\rm M}_{f,\delta_n}(X_n,\mathcal I_n), {\rm R}_f(\mathbb X)) \leq d_c) \geq c$. This distance can then be interpreted directly on the persistence
diagrams by drawing boxes of radius $d_c$ around each point. If the box of a point intersects the diagonal, it means that we cannot guarantee that the corresponding feature 
will not have size 0, i.e. will disappear,
in the target ${\rm R}_f(\mathbb X)$. On the other hand, an empty intersection with the diagonal ensures that the feature is meaningful at level $c$.



\subsection{Extension to multivariate Mappers}
\label{sec:multiMap}

In this section, we extend the previous results to the case where the Mappers are computed with multivariate functions $f:X\rightarrow\R^p$.
Let us first define the Mapper computed in practice with a multivariate filter.

\begin{defin}
Let $X\subset\R^d$ be a point cloud, and $f:X\rightarrow\R^p$ be a multivariate function defined on $X$. Let 
$\mathcal U=\{U_i\}$ be a hypercube cover of ${\rm im}(f)$, i.e. a family of hypercubes such that ${\rm im}(f)\subseteq \cup_i U_i$, 
let $\mathcal V = f^{-1}(\mathcal U) = \{f^{-1}(U_i)\}$
be the cover obtained by taking the preimage of $\mathcal I$ under $f$.
Let $\tilde{\mathcal V} = {\rm SL}_\delta(\mathcal V)$ be the cover obtained with
single-linkage clustering with parameter $\delta$ on each element of $\mathcal V$. 
Then, the {\em Mapper} ${\rm M}_{f,\delta}(X,\mathcal U)$ is defined as:
\begin{equation}
{\rm M}_{f,\delta}(X,\mathcal U) = \mathcal N (\tilde{\mathcal V}),
\end{equation} 
where $\mathcal N$ denotes the nerve operation.
\end{defin}

Note that the choice of $\delta_n$ is independent from the filters, so we can safely use the same $\delta_n$ as in the previous section, at least
for the study of 0- and 1-dimensional homology.
The following result is a straightforward extension of Theorem~\ref{th:1dmap}. 

\begin{thm}\label{thm:paramselectmulti}
Let $X_n\subset\R^d$ be a point cloud, and 
let $f:X_n\rightarrow\R^p$ be a function defined on $X_n$.
Let $\delta_n=d_{\rm H}(X_n,X_{s(n)})$, where $s(n)$ is defined as in Equation~(\ref{eq:sn}).
Let $\mathcal U_n$ be a hypercube cover of ${\rm im}(f)$ such that, for all $1\leq s \leq p$, the $s$-sides of all hypercubes have gain $g_s\in\left(\frac 13, \frac 12\right)$ 
and resolution $r_s = {\rm max}\{|f_s(x)-f_s(y)|\,:\,\|x-y\|\leq\delta_n\} / g_s$, where $f_s$ denotes the $s$-th coordinate of $f$ and the $s$-side of the 
hypercube $I_1\times\cdots\times I_p$ is the interval $I_s$.
Then ${\rm M}_{f,\delta_n}(X_n,\mathcal U_n)$ and ${\rm M}_f(G_{\delta_n}(X_n),\mathcal U_n)$ are isomorphic.
\end{thm}

Finally, we also extend the bottleneck distance to multivariate Mappers by simply aggregating distances in all dimensions:

\begin{defin}
Let ${\rm M},{\rm M}'$ be two multivariate Mappers. 
Then, the {\em multivariate bottleneck distance} is defined as:
\begin{equation}
d_{\rm b}({\rm M},{\rm M}') = {\rm max}\{d^s_{\rm b}({\rm M}, {\rm M}')\,:\, 1\leq s \leq p\},
\end{equation}
where $d^s_{\rm b}$ denotes the bottleneck distance between the extended persistence diagrams of the Mappers computed with the $s$-th coordinate of their filters. 
\end{defin}

Using bootstrapping with this metric allows us, as in the previous section, to derive the confidence level at which all topological features of a certain size in the multivariate Mapper are preserved,
or to visualize the features that are guaranteed at a given confidence level. Moreover, we expect this extended distance to also satisfy stability 
and convergence properties in 0- and 1-dimensional homology, since the proofs of these properties in the case where the filter is scalar-valued 
that are presented in~\cite{Carriere17b} and~\cite{Carriere18a} should extend almost straightforwardly. 
We are now ready to apply the Mapper on Hi-C datasets.

\section{Application to single-cell Hi-C contact maps}

In this section, we study the dataset of~\cite{Nagano17}. In this article, the authors generated thousands of fragment pairs from the 21 chromosomes of each of 1171
${\rm F}_1$ hybrid 129 $\times$ Castaneus mouse ES cells, and showed that several features of this distribution of pairs, such as the ratio between long and short contacts, were directly correlated
with the cell cycle phases. Moreover, this dataset was also studied in~\cite{Liu18}, in which the authors demonstrated how the SCC could be used to
embed cells in a lower-dimensional space in which the cell cycle phases are clearly separated along a circular shape. In this section, 
we demonstrate how the biological factors correlated with the cell cycle phases described in~\cite{Nagano17} can be validated on the Mapper, and we show
how the Mapper can be used to formally retrieve this circular shape and to compute confidence levels on it. 

\subsection{Method}

\begin{figure}\centering
\includegraphics[width = 7.5cm]{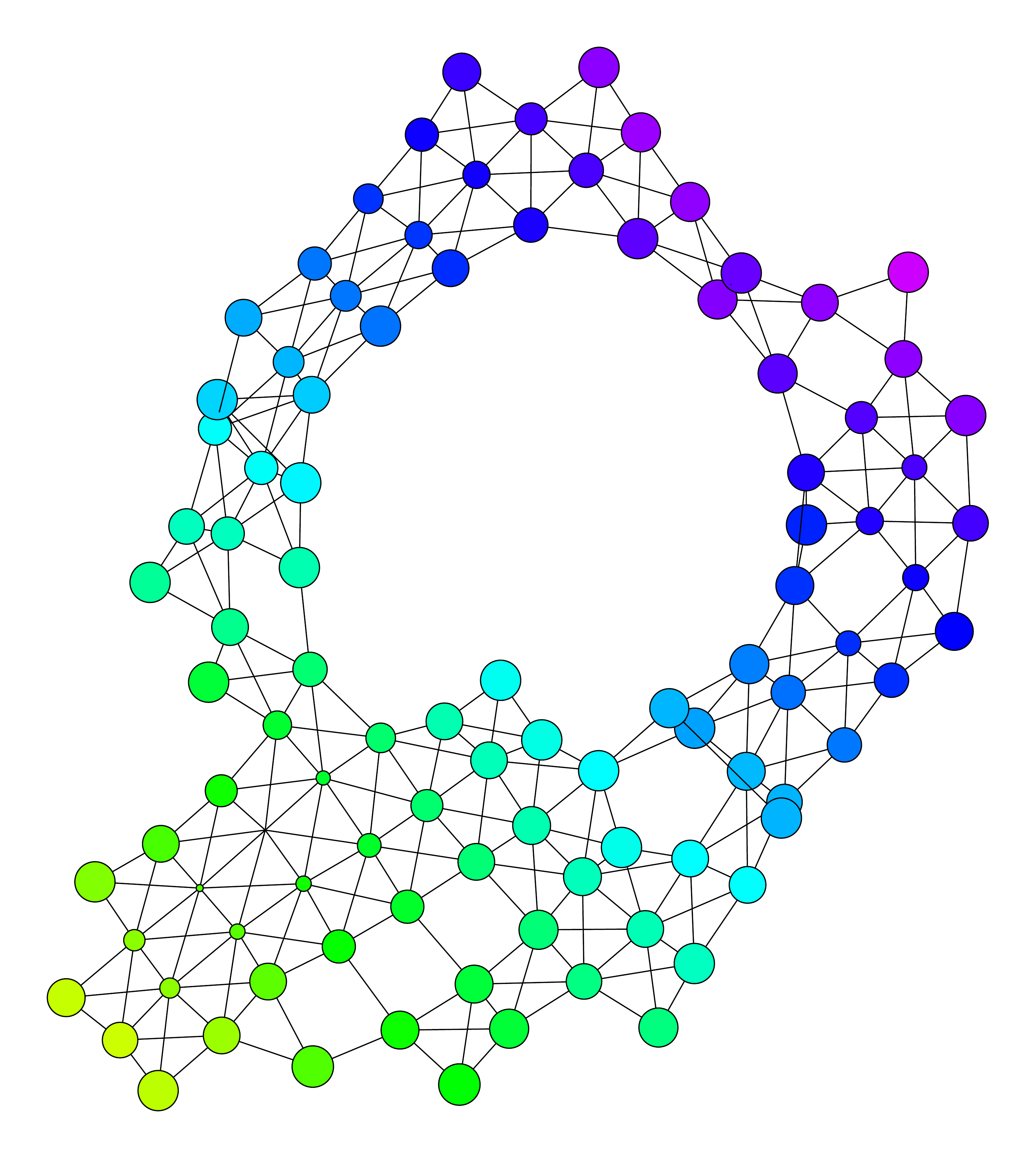}
\includegraphics[width = 7.5cm]{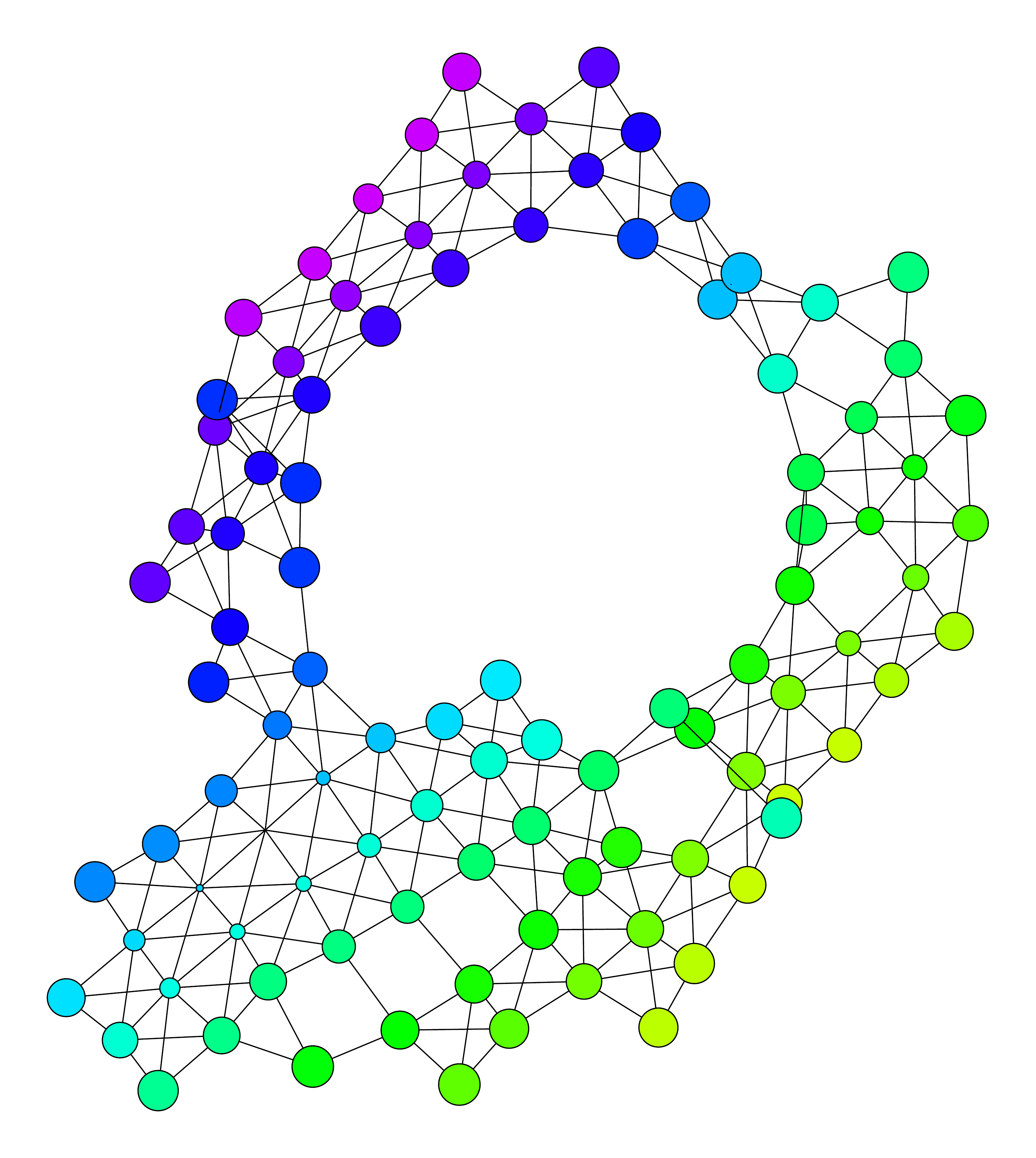}
\includegraphics[width = 7.5cm]{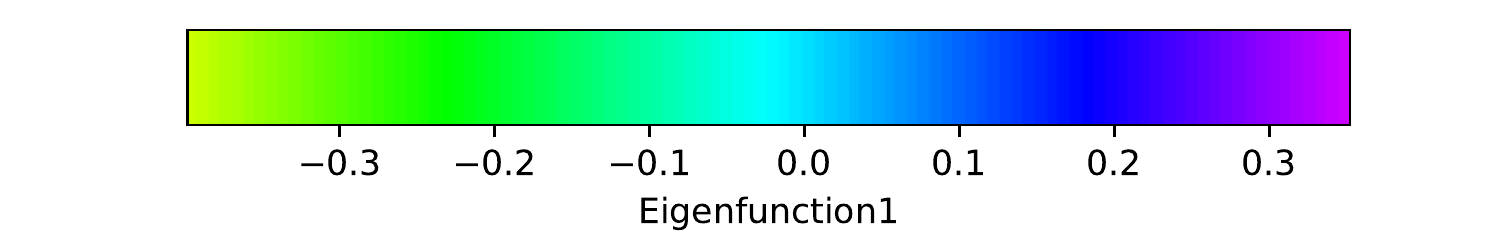}
\includegraphics[width = 7.5cm]{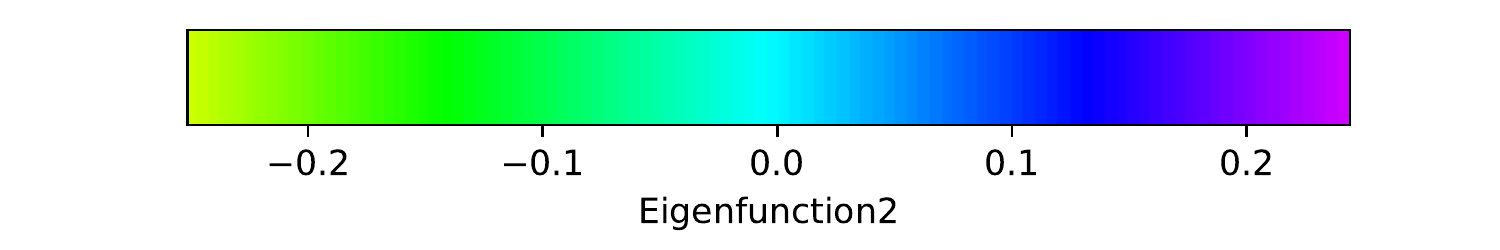}
\includegraphics[width = 7.5cm]{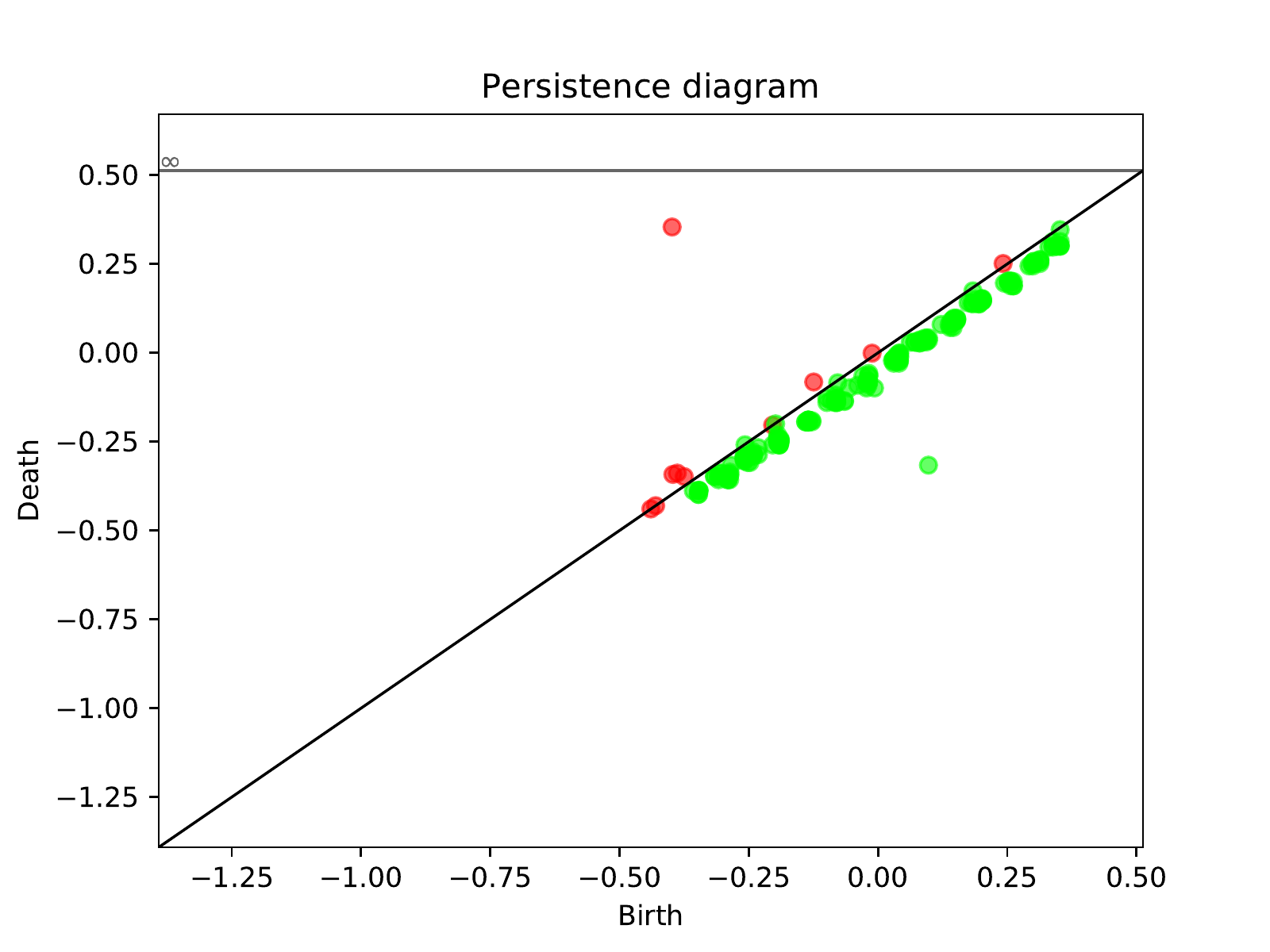}
\includegraphics[width = 7.5cm]{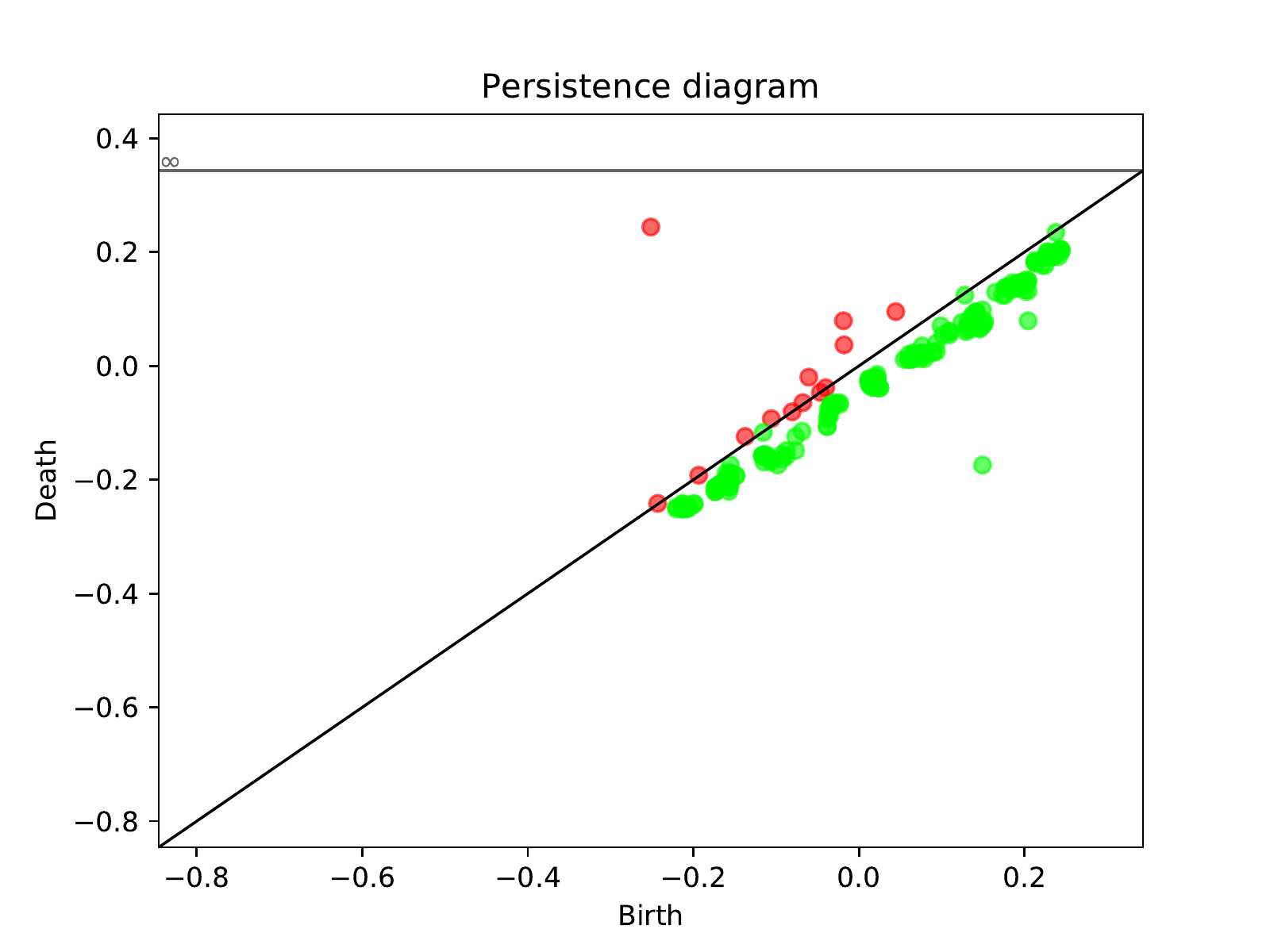}
\caption{\label{fig:mapeig1} Mapper graph colored with the first (left column) and second (right column) eigenfunctions. We also show the corresponding 
persistence diagrams used to characterize the topology of the graphs. Red points correspond to topological features in dimension 0 such as connected components, while green points
correspond to topological features in dimension 1, such as loops.}
\end{figure}

We first turned the fragment pairs into contact matrices using 500kb resolution bins. We smoothed the matrices with a moving average window of size 1, as recommended in~\cite{Liu18}, and then
computed all pairwise SCC values between contact matrices to generate a 1171 $\times$ 1171 similarity matrix. 
A corresponding distance matrix was derived from the similarities by using the relation: $$d_{\rm SCC}(X,Y) = \sqrt{{\rm SCC}(X,X) + {\rm SCC}(Y,Y) - 2{\rm SCC}(X,Y)},$$
and we finally computed the Mapper from this distance matrix.
The Mapper filters used were the first two eigenfunctions of the dataset, and the Mapper parameters 
were computed automatically as stated in Theorem~\ref{thm:paramselectmulti}. 
Moreover, we restricted to 0- and 1-dimensional topology to ease visualization and interpretation. This means in particular that we only observed the 1-skeleton
of the Mapper and did not consider the higher-dimensional simplices.
The obtained Mapper, colored by the first two eigenvalues, is displayed in Figure~\ref{fig:mapeig1}. 

\subsection{Biological interpretation}

\begin{figure}\centering
\includegraphics[width = 8cm]{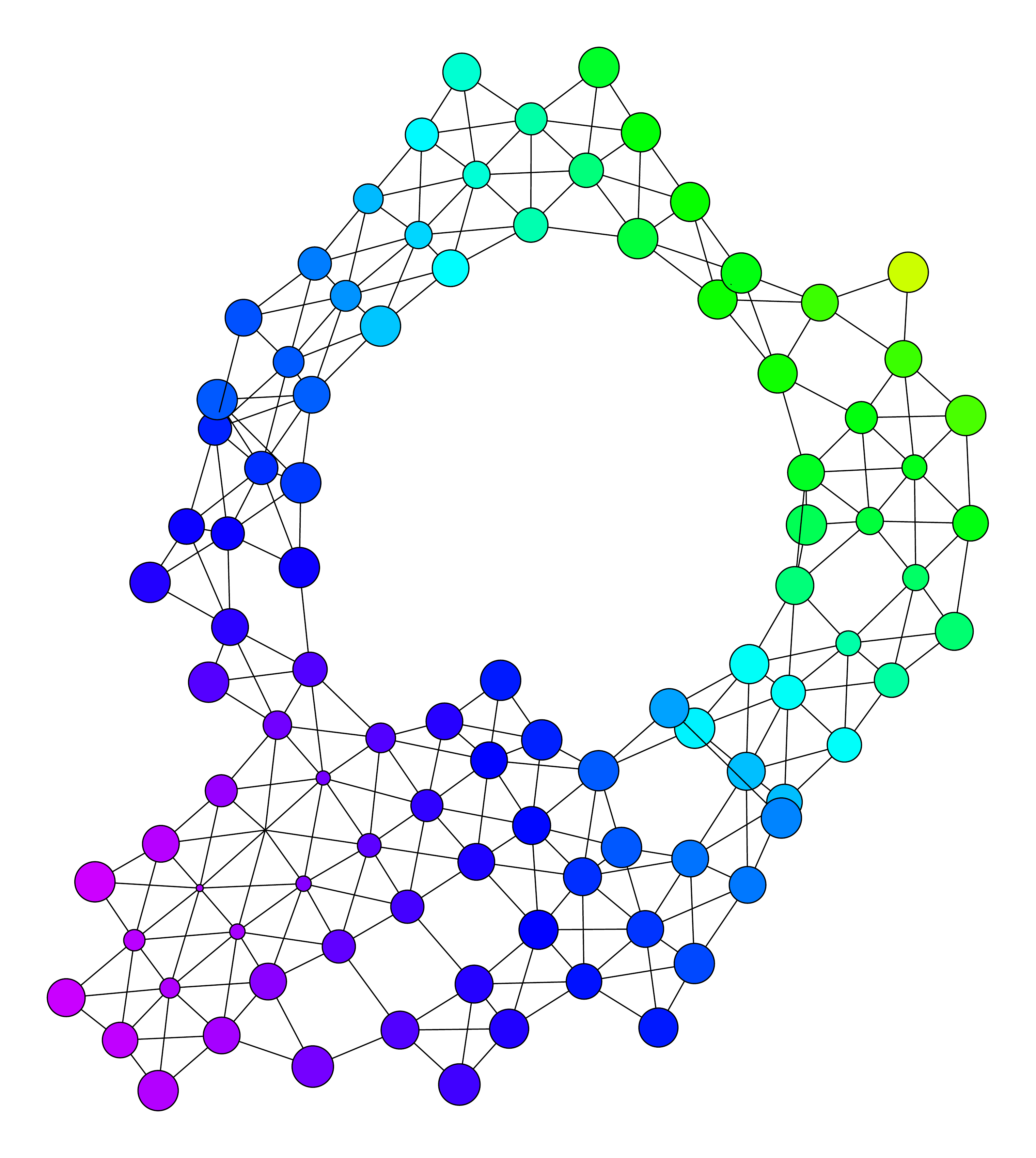}
\includegraphics[width = 8cm]{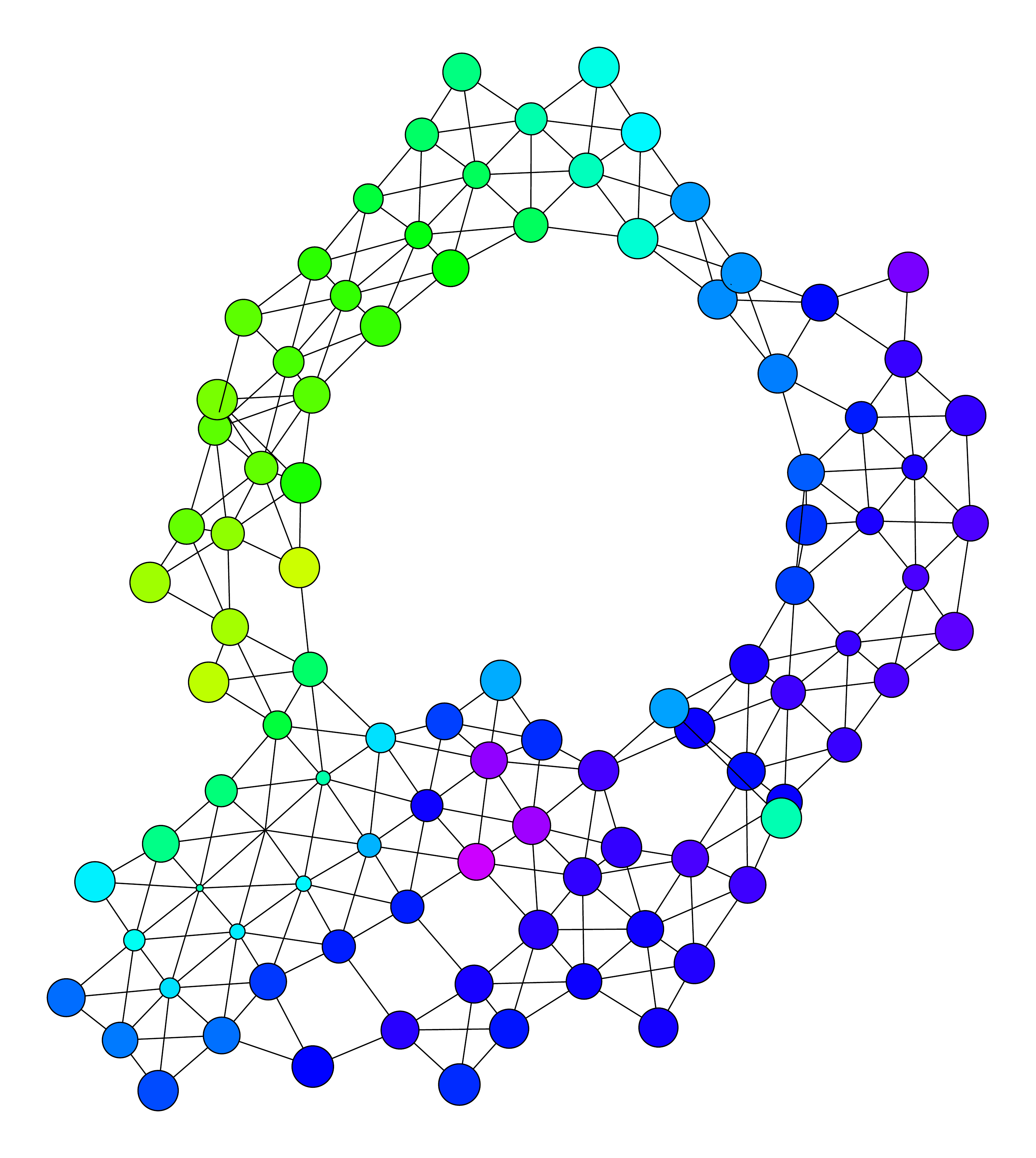}
\includegraphics[width = 7.5cm]{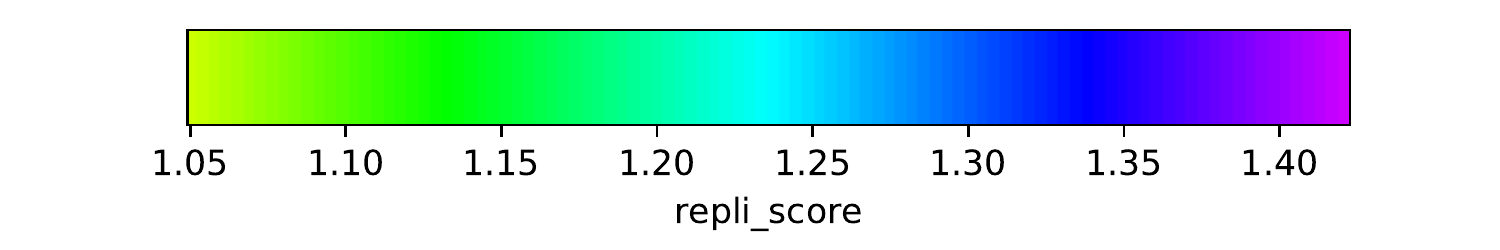}
\includegraphics[width = 7.5cm]{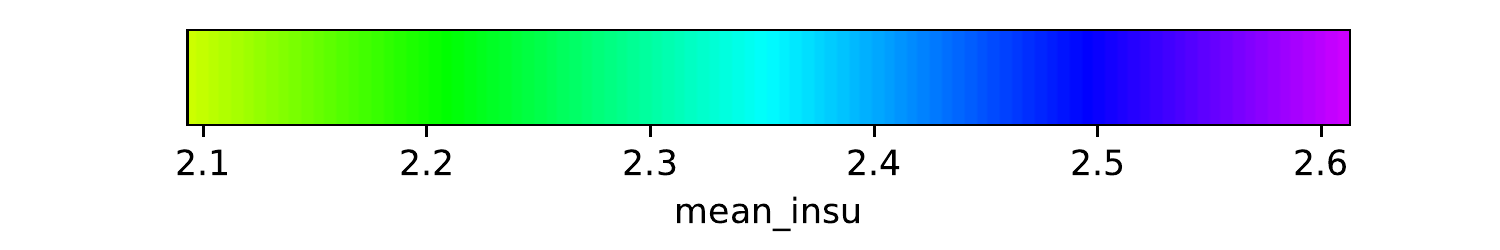}
\includegraphics[width = 8cm]{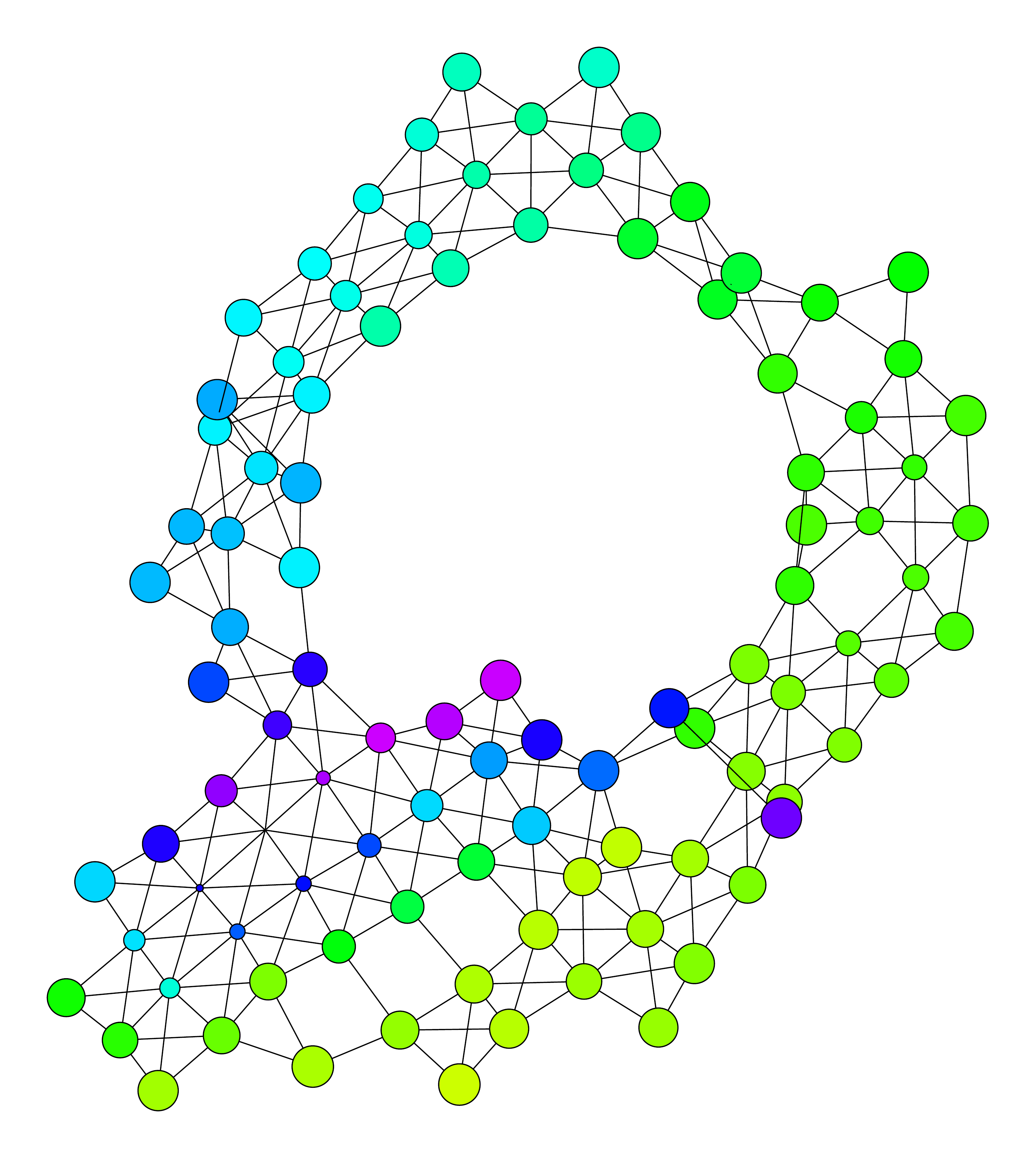}
\includegraphics[width = 8cm]{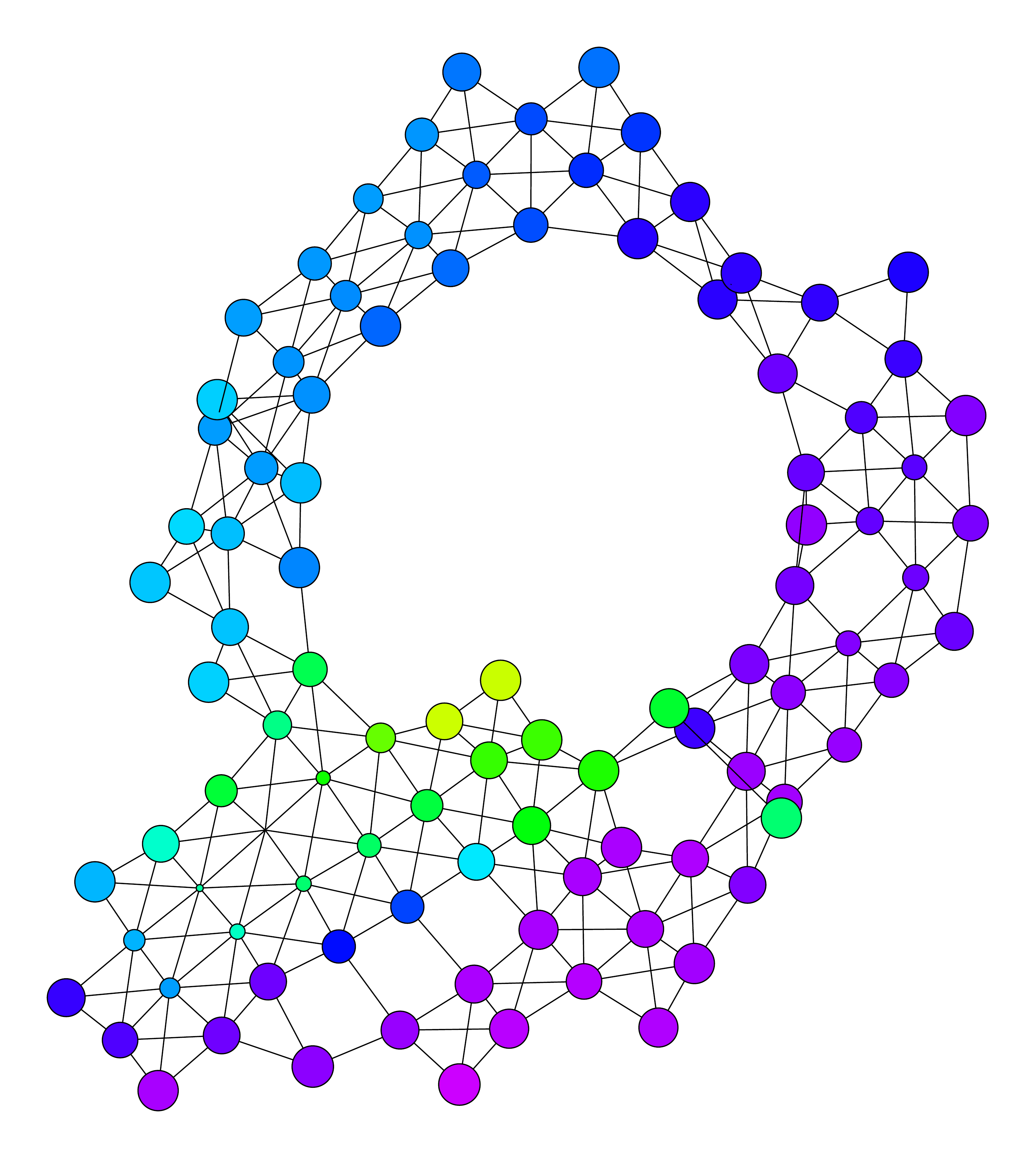}
\includegraphics[width = 7.5cm]{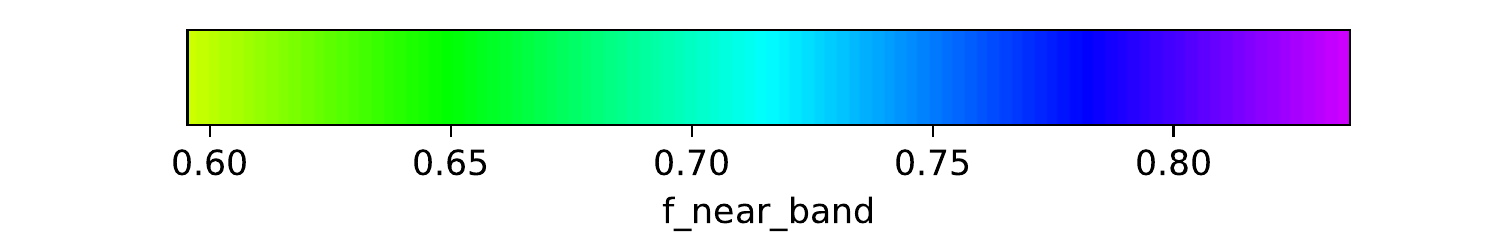}
\includegraphics[width = 7.5cm]{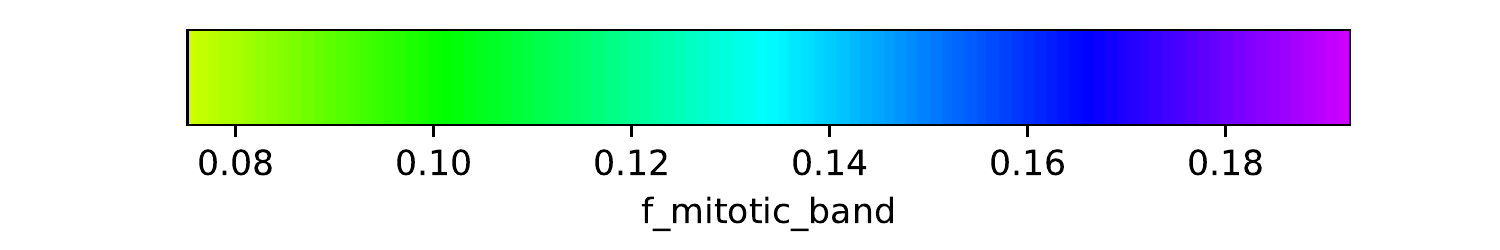}
\caption{\label{fig:mapbio} Mapper graphs colored with the biological factors correlated with the cell cycle phases as observed in~\cite{Nagano17}. For each biological marker,
the distribution of values is consistent with the empirical observations made by the authors.}
\end{figure}

As one can see on the graphs, a cycle is clearly visible, which probably represents the cell cycle.
A useful property of the Mapper is its ability to encode and visualize correlations among multiple variables. For this dataset, the authors in~\cite{Nagano17}
demonstrated that several biological quantities were associated with the cell cycle progression. In particular, they showed that the so-called
repli-score, which is roughly the ratio between the copy-number of genomic regions associated with early-replicating phases of the cell cycle and the total number of reads
(see the exact definition of this score in~\cite{Nagano17}) was highly correlated with cell cycle progression, which is clearly visible 
on the Mapper as well: if we color the Mapper nodes with the repli score values (see upper left corner of Figure~\ref{fig:mapbio}),
one can easily see that the values gradually increase and decrease along the cycle. The authors also noticed that the mean insulation of topologically associated domain borders
was another marker which was highly correlated with the cell cycle, which, again, can be retrieved from the Mapper colored by this insulation (see upper right corner of
Figure~\ref{fig:mapbio}). 

Another interesting observation in~\cite{Nagano17} was that the percentage of long-range distances 
(between 2 and 12 Mb, "mitotic band") and the one of short-range distances (less than 2 Mb, "near band") of the contact matrices were characteristic of specific 
phases of the cell cycle. In particular,due to the highly condensed structure of the chromosomes during mitosis, cells belonging to this phase tended to have 
more long-range distances, while the opposite was observed for cells exiting mitosis, since chromosomes decondense in preparation for DNA transcription. 
Plotting either the short-range percentage (lower-left corner of Figure~\ref{fig:mapbio}) or the long-range one (lower-right corner of Figure~\ref{fig:mapbio}) 
allows us to validate this observation, since it is clearly visible from the graphs that the cycle can be divided into regions with distinct distribution values 
that correspond to the cell cycle phases. 

\subsection{Formal encoding and statistical significance}

\begin{figure}[h!]\centering
\includegraphics[width = 7.5cm]{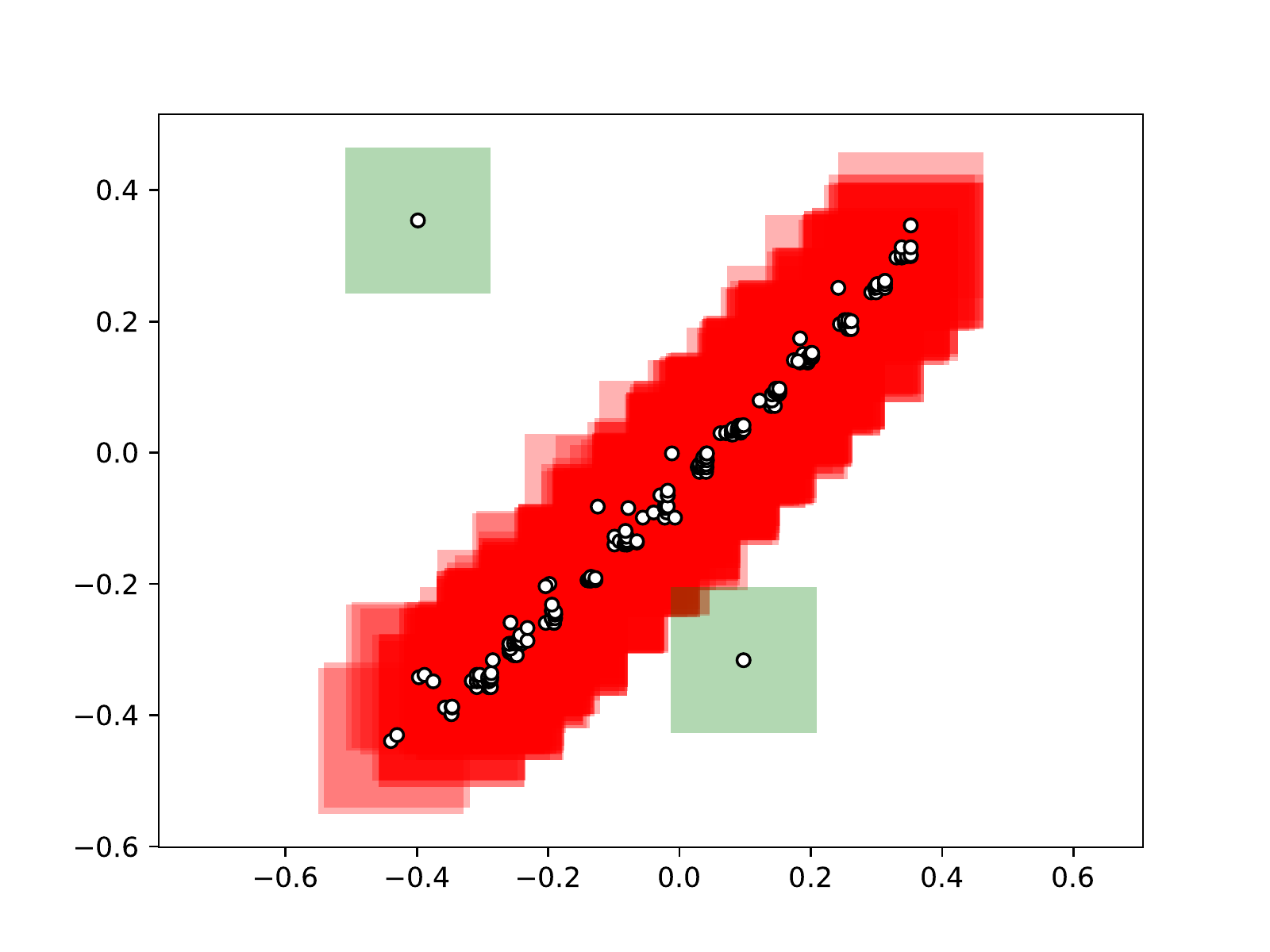}
\includegraphics[width = 7.5cm]{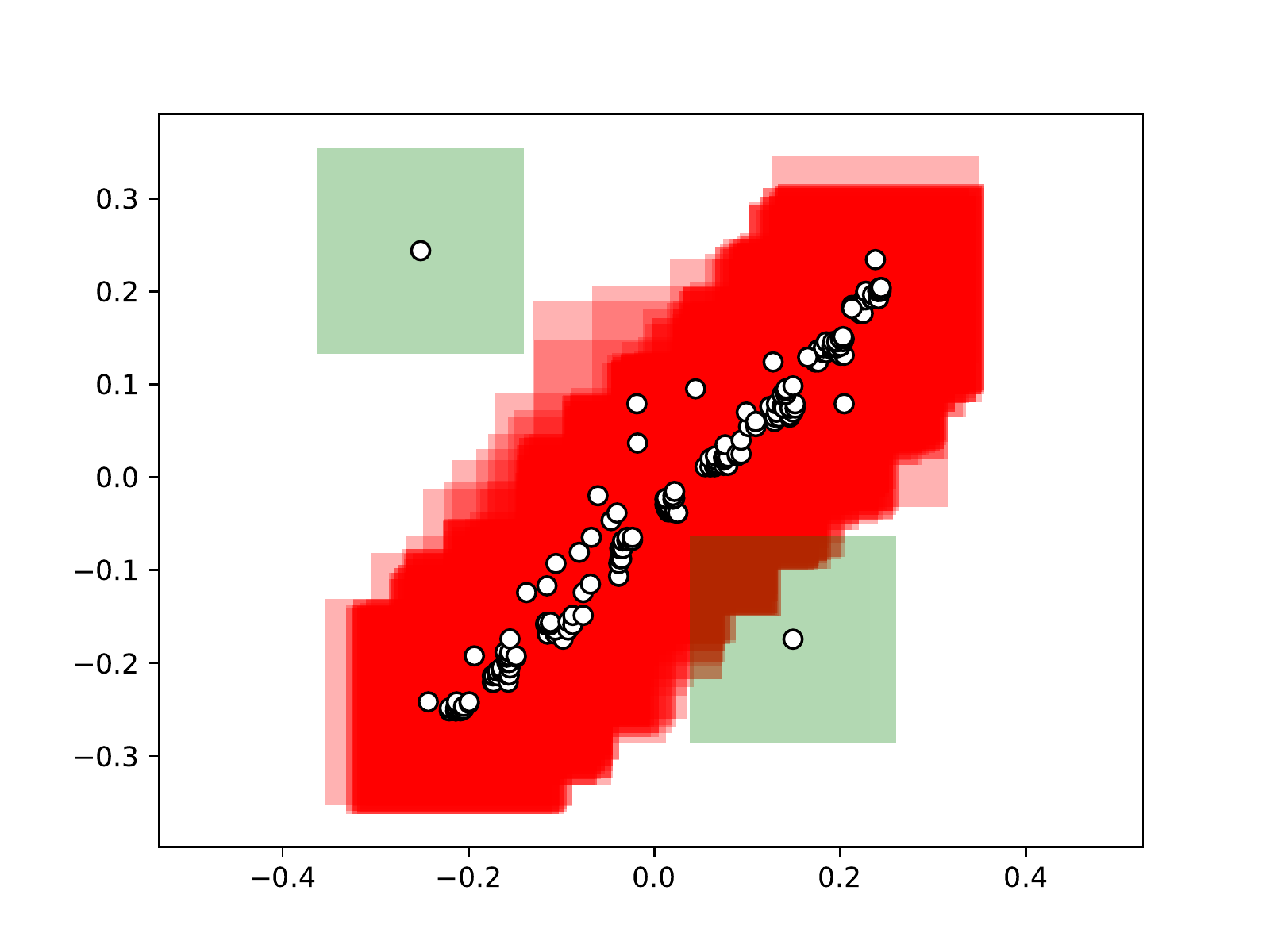}
\caption{\label{fig:conf90} Confidence region at level 90\% retrieved from bootstrapping the Mapper through its set of extended persistence diagrams. Confidence boxes
are drawn around each point, and colored according to whether they intersect the diagonal (red) or not (green).}
\end{figure}

We recall that we showed in Section~\ref{sec:stats} that extended persistence diagrams can be used to formally assess the topological features of Mappers.
In this dataset, the presence of the cell cycle can be noted from the extended persistence diagrams of the first two eigenfunctions, since there is a green point
which clearly stands out from the others. In order to statistically validate the presence of this loop, we apply our method defined in Section~\ref{sec:multiMap}, 
and we used $100$-fold bootstrapping to generate a distance distribution. We show the corresponding 90\% confidence region on the diagrams of the 
two first eigenvalues in Figure~\ref{fig:conf90}. In both diagrams, the two points corresponding to the connected component and the loop of the graph have 
confidence boxes that do not intersect the diagonal, which means their confidence is at least 90\%. 
In fact, the confidence level computed for the point corresponding to the loop was around 93\%, meaning that there is a strong confidence that the cell cycle that we retrieved on
the data is relevant and not due to noise or artifacts.     

\section{Conclusion}

In this article, we provided a mathematical way to process datasets of contact maps with statistical guarantees by making use of Topological Data Analysis. 
We demonstrated that combining the statistical power and theoretical formalization of the Mapper algorithm with the stratum-adjusted correlation coefficient enabled the 
validation of biological factors responsible for the topological structure, and the quantification of the confidence one can have for it. As for future directions, we plan to
theoretically study the statistical guarantees and properties of our bootstrap extension to multivariate Mapper and to further investigate 
the generalizability of the application of TDA to contact matrices by studying more datasets from the literature, and seeing if the 
underlying biological processes can also be retrieved in the corresponding Mapper.

\section*{Acknowledgements}
This work has been funded by NIH grants (U54 CA193313 and U54 CA209997) and Chan Zuckerberg Initiative pilot grant.

\newpage

\bibliographystyle{alpha}
\bibliography{biblio}

\newcommand{\etalchar}[1]{$^{#1}$}
\begin{thebibliography}{LAvBW{\etalchar{+}}09}

\bibitem[Agr12]{Agresti12}
Alan Agresti.
\newblock {\em Categorical data analysis, 3rd edition}.
\newblock Wiley, 2012.

\bibitem[AN15]{Ay15}
Ferhat Ay and William Noble.
\newblock {Analysis methods for studying the 3D architecture of the genome}.
\newblock {\em Genome Biology}, 16:183--198, 2015.

\bibitem[Car09]{Carlsson09a}
Gunnar Carlsson.
\newblock Topology and data.
\newblock {\em Bulletin of the American Mathematical Society}, 46:255--308,
  2009.

\bibitem[CLR16]{Camara16}
Pablo Camara, Arnold Levine, and Raul Rabadan.
\newblock {Inference of Ancestral Recombination Graphs through Topological Data
  Analysis}.
\newblock {\em PLoS Computational Biology}, 12(8):1--25, 2016.

\bibitem[CMO18]{Carriere18a}
Mathieu Carri{\`e}re, Bertrand Michel, and Steve Oudot.
\newblock Statistical analysis and parameter selection for mapper.
\newblock {\em Journal of Machine Learning Research}, 19(12):1--39, 2018.

\bibitem[CO17]{Carriere17b}
Mathieu Carri\`ere and Steve Oudot.
\newblock {Structure and Stability of the 1-Dimensional Mapper}.
\newblock {\em Foundations of Computational Mathematics}, 2017.

\bibitem[CSEH07]{Cohen07}
David Cohen-Steiner, Herbert Edelsbrunner, and John Harer.
\newblock {Stability of Persistence Diagrams}.
\newblock {\em Discrete and Computational Geometry}, 37(1):103--120, 2007.

\bibitem[CSEH09]{Cohen09}
David Cohen-Steiner, Herbert Edelsbrunner, and John Harer.
\newblock {Extending persistence using Poincar{\'e} and Lefschetz duality}.
\newblock {\em Foundation of Computational Mathematics}, 9(1):79--103, 2009.

\bibitem[DMW16]{Dey16}
Tamal Dey, Facundo M{\'{e}}moli, and Yusu Wang.
\newblock {Multiscale Mapper: Topological Summarization via Codomain Covers}.
\newblock In {\em Proceedings of the 27th Symposium on Discrete Algorithms},
  pages 997--1013, 2016.

\bibitem[DRA{\etalchar{+}}06]{Dostie06}
Jos\'ee Dostie, Todd Richmond, Ramy Arnaout, Rebecca Selzer, William Lee,
  Tracey Honan, Eric Rubio, Anton Krumm, Justin Lamb, Chad Nusbaum, Roland
  Green, and Job Dekker.
\newblock {Chromosome Conformation Capture Carbon Copy (5C): A massively
  parallel solution for mapping interactions between genomic elements}.
\newblock {\em Genome Research}, 16(10):1299--1309, 2006.

\bibitem[DRDK02]{Dekker02}
Job Dekker, Karsten Rippe, Martijn Dekker, and Nancy Kleckner.
\newblock Capturing chromosome conformation.
\newblock {\em Science}, 295(5558):1306--1311, 2002.

\bibitem[dWdL12]{Wit12}
Elzo de~Wit and Wouter de~Laat.
\newblock {A decade of 3C technologies: insights into nuclear organization}.
\newblock {\em Genes and Development}, 26(1):11--24, 2012.

\bibitem[EH10]{Edelsbrunner10}
Herbert Edelsbrunner and John Harer.
\newblock {\em {Computational Topology: an introduction}}.
\newblock AMS Bookstore, 2010.

\bibitem[LAvBW{\etalchar{+}}09]{Aiden09}
Erez Lieberman-Aiden, Nynke van Berkum, Louise Williams, Maxim Imakaev, Tobias
  Ragoczy, Agnes Telling, Ido Amit, Bryan Lajoie, Peter Sabo, Michael
  Dorschner, Richard Sandstrom, Bradley Bernstein, Michael Bender, Mark
  Groudine, Andreas Gnirke, John Stamatoyannopoulos, Leonid Mirny, Eric Lander,
  and Job Dekker.
\newblock {Comprehensive Mapping of Long-Range Interactions Reveals Folding
  Principles of the Human Genome}.
\newblock {\em Science}, 326(5950):289--293, 2009.

\bibitem[LLYN18]{Liu18}
Jie Liu, Dejun Lin, Galip Yardimci, and William Noble.
\newblock {Unsupervised embedding of single-cell Hi-C data}.
\newblock {\em Bioinformatics}, 34(13):i96--i104, 2018.

\bibitem[Man63]{Mantel63}
Nathan Mantel.
\newblock {Chi-Square Tests with One Degree of Freedom; Extensions of the
  Mantel-Haenszel Procedure}.
\newblock {\em Journal of the American Statistical Association},
  58(303):690--700, 1963.

\bibitem[MW16]{Munch16}
Elizabeth Munch and Bei Wang.
\newblock {Convergence between Categorical Representations of Reeb Space and
  Mapper}.
\newblock In {\em Proceedings of the 32nd Symposium on Computational Geometry},
  volume~51, pages 53:1--53:16, 2016.

\bibitem[NLV{\etalchar{+}}17]{Nagano17}
Takashi Nagano, Yaniv Lubling, Csilla Varnai, Carmel Dudley, Wing Leung, Yael
  Baran, Netta Cohen, Steven Wingett, Peter Fraser, and Amos Tanay.
\newblock {Cell-cycle dynamics of chromosomal organization at single-cell
  resolution}.
\newblock {\em Nature}, 547:61--67, 2017.

\bibitem[Oud15]{Oudot15}
Steve Oudot.
\newblock {\em {Persistence Theory: From Quiver Representations to Data
  Analysis}}.
\newblock Number 209 in Mathematical Surveys and Monographs. {American
  Mathematical Society}, 2015.

\bibitem[RCK{\etalchar{+}}17]{Rizvi17}
Abbas Rizvi, Pablo Camara, Elena Kandror, Thomas Roberts, Ira Schieren, Tom
  Maniatis, and Raul Rabadan.
\newblock {Single-cell topological RNA-seq analysis reveals insights into
  cellular differentiation and development}.
\newblock {\em Nature Biotechnology}, 35:551--560, 2017.

\bibitem[SKS{\etalchar{+}}06]{Simonis06}
Marieke Simonis, Petra Klous, Erik Splinter, Yuri Moshkin, Rob Willemsen, Elzo
  de~Wit, Bas van Steensel, and Wouter de~Laat.
\newblock {Nuclear organization of active and inactive chromatin domains
  uncovered by chromosome conformation capture-on-chip (4C)}.
\newblock {\em Nature Genetics}, 38:1348--1354, 2006.

\bibitem[SMC07]{Singh07}
Gurjeet Singh, Facundo M\'emoli, and Gunnar Carlsson.
\newblock {Topological Methods for the Analysis of High Dimensional Data Sets
  and 3D Object Recognition}.
\newblock In {\em Symposium on Point Based Graphics}, pages 91--100, 2007.

\bibitem[YZY{\etalchar{+}}17]{Yang17}
Tao Yang, Feipeng Zhang, Galip Yardimci, Fan Song, Ross Hardison, William
  Noble, Feng Yue, and Qunhua Li.
\newblock {HiCRep: assessing the reproducibility of Hi-C data using a
  stratum-adjusted correlation coefficient}.
\newblock {\em Genome Research}, 27(11):1939--1949, 2017.

\end{thebibliography}

\end{document}